\documentclass[11pt]{article}
\usepackage{amsmath,amssymb,color,graphics,epsfig,cite}

\usepackage{amsfonts}

\newcommand{\hoch}[1]{$\, ^{#1}$}

\newcommand{\be}{\begin{equation}}
\newcommand{\ee}{\end{equation}}
\newcommand{\bea}{\setlength\arraycolsep{2pt} \begin{eqnarray}}
\newcommand{\eea}{\end{eqnarray}}
\newcommand{\nn}{\nonumber}

\def\fft#1#2{{\frac{#1}{#2}}}

\def\0{{\sst{(0)}}}
\def\1{{\sst{(1)}}}
\def\2{{\sst{(2)}}}
\def\3{{\sst{(3)}}}
\def\4{{\sst{(4)}}}
\def\5{{\sst{(5)}}}
\def\6{{\sst{(6)}}}
\def\7{{\sst{(7)}}}
\def\8{{\sst{(8)}}}
\def\sst#1{{\scriptscriptstyle #1}}

\begin{document}

\begin{center}
{\Large {\bf A Space/Time Interchange Symmetry\\ of Rotating AdS Black Holes in General Dimensions}}

\vspace{20pt}

Si-Yue Lu\hoch{1}, Peng Zhao\hoch{1} and H. L\"{u}\hoch{2,1}

\vspace{10pt}

{\it \hoch{1}Joint School of the National University of Singapore and Tianjin University,\\
International Campus of Tianjin University, Binhai New City, Fuzhou 350207, China}

\bigskip

{\it \hoch{2}Center for Joint Quantum Studies and Department of Physics,\\
School of Science, Tianjin University, Tianjin 300350, China }

\vspace{40pt}

\underline{ABSTRACT}
\end{center}

We revisit the previously known local inversion symmetry of the five-dimensional Kerr-AdS metric that relates the over-rotating black hole to the under-rotating one and reinterpret it as an interchanging symmetry between time and the longitudinal angular coordinates. We generalize this to all $D$ dimensions, including $D=4$, thereby enlarging the trivial linear $\mathbb Z_N$ symmetry of the $N=\lfloor(D-1)/2\rfloor$ longitudinal angular coordinates to the nonlinearly realized $\mathbb Z_{N+1}$ symmetry that involves time.

\vfill{siyue\_lu@tju.edu.cn\ \ \ pzhao@tjufz.org.cn\ \ \ mrhonglu@gmail.com}


\thispagestyle{empty}
\pagebreak


\newpage

\section{Introduction}

Although time and space are united in both Einstein's Special and General Relativity, they remain distinct concepts. In Special Relativity, time and space can ``rotate'' non-compactly via the Lorentz boost, but unlike two space coordinates, they cannot interchange. In General Relativity, the roles of time and space coordinates may swap while crossing the null hyper surface, but there is no symmetry between them. In this paper, we propose a new discrete local symmetry that interchanges the time coordinate with any one of the Killing directions in the foliating sphere in the rotating anti-de Sitter (AdS) spacetime in general dimensions.

The Einstein metrics of rotating AdS black holes in diverse dimensions were constructed in \cite{Gibbons:2004uw,Gibbons:2004js}, which generalized ($D=4$) Carter \cite{Carter:1968ks} and ($D=5$) Hawking-Hunter-Taylor \cite{Hawking:1998kw} solutions. The further generalization to higher-dimensional Plebanski metric with Taub-NUT parameters was given in \cite{Chen:2006ea,Chen:2006xh}, where it was observed that the Taub-NUT parameter is trivial in five dimensions since it can be removed by a simple coordinate transformation. Furthermore, a discrete {\it inversion} symmetry was observed in five dimensions that relates over-rotating black holes to under-rotating ones. Specifically, in $D$ dimensions, the foliating $(D-2)$-sphere has $N=\lfloor(D-1)/2\rfloor$ orthogonal Killing directions $\partial/\partial \phi_i$, supporting $N$ independent rotations. Typically, one uses constants $a_i$ to parameterize these rotations. For AdS black holes, one requires that $\Xi_i=1 - a_i^2 g^2\ge 0$, where $g=1/\ell$ is the inverse of the AdS radius. Metrics with $\Xi_i<0$ describe over-rotating geometries. It was observed that in five dimensions, there exists a remarkable inversion symmetry, namely the ill-defined over-rotating geometry is locally related to the well-defined under-rotating black hole through a coordinate transformation. In particular, for the Hawking-Hunter-Taylor metric, the transformation of the three commuting Killing directions are \cite{Chen:2006ea}
\be
\phi\rightarrow - \fft{1}{ag} \phi\,,\qquad
\psi \rightarrow \psi - \fft{b}{a}\phi\,,\qquad t\rightarrow a g t +\fft{1}{g} \phi\,.
\ee
The physical meaning of the above transformation is obscure and there has been no further development. We note that the original Hawking-Hunter-Taylor metric is asymptotically rotating with angular velocities $\Omega_{a}^\infty=-a g^2$ and $\Omega_{b}^\infty=-b g^2$. If we write it in the asymptotically non-rotating frame, the transformation among the Killing directions is simply
\be
\phi\leftrightarrow g t\,,\label{interchange}
\ee
i.e.~it is a symmetry that interchanges space and time.

This paper aims to generalize this five-dimensional discrete local symmetry to rotating black holes in general dimensions. The paper is organized as follows. In section 2, we briefly review the previously known inversion symmetry and then translate it to the space/time interchange symmetry. In section 3, we generalize the symmetry to $D$ dimensions but pay special attention to $D=4$ and 3. We study the symmetry's effect on the mass and angular momentum in these two dimensions. We conclude the paper in section 4.

\section{Improved symmetry transformation rules in $D=5$}

In this section, we review the original local inversion symmetry which was found \cite{Chen:2006ea} in five-dimensional Kerr-AdS black hole metric and then improve the transformation rules by writing the metric in the non-rotating frame in the asymptotic region. The five-dimensional Kerr-AdS metric was first constructed by Hawking, Hunter and Taylor. The original metric is \cite{Hawking:1998kw}
\bea
ds^2 &=& -\frac{\Delta_r}{\rho^2}\left(dt-\frac{a \sin^2\theta}{\Xi_a}d\phi-\frac{b \cos^2\theta}{\Xi_b}d\psi\right)^2+\frac{\Delta_\theta \sin^2\theta}{\rho^2}\left(a dt - \frac{r^2+a^2}{\Xi_a}d\phi\right)^2 \nn\\
&& +\frac{\Delta_\theta \cos^2\theta}{\rho^2}\left(bdt-\frac{r^2+b^2}{\Xi_b}d\psi\right)^2 +\frac{\rho^2 dr^2}{\Delta_r}+\frac{\rho^2 d\theta^2}{\Delta_\theta}\nn\\
&& +\frac{1+r^2g^2}{r^2 \rho^2}\left(a bdt-\frac{b(r^2+a^2)\sin^2\theta}{\Xi_a}d\phi-
\frac{a(r^2+b^2)\cos^2\theta}{\Xi_b}d\psi\right)^2,\label{hht}
\eea
where
\bea
\Delta_r &=& \frac{1}{r^2}(1+r^2g^2)(r^2+a^2)(r^2+b^2)-2m\,,\qquad
\Delta_\theta = 1-a^2g^2\cos^2\theta-b^2g^2\sin^2\theta\,,\nn\\
\rho^2 &=& r^2 + a^2\cos^2\theta + b^2\sin^2\theta\,,\qquad
\Xi_a = 1- a^2g^2\,,\qquad  \Xi_b = 1 -b^2g^2\,.
\eea
This metric is invariant under inversion of the rotation parameter $a$ or $b$, followed by an appropriate coordinate transformation \cite{Chen:2006ea}
\bea
m&=&\frac{\tilde{m}}{(\tilde{a} g)^4}\,,\qquad a=\frac{1}{\tilde{a}g^2}\,,\qquad b=\frac{\tilde{b}}{\tilde{a}g}\,,\nn\\
r&=&\frac{1}{\tilde{a}g}\tilde{r}\,,\qquad \cos\theta=\left(1-\frac{\tilde{\Xi}_a}{\tilde{\Xi}_b}\right)^{\frac{1}{2}}
\cos\tilde{\theta}\,,\nn\\
\phi&=&-\frac{1}{\tilde{a}g}\tilde{\phi}\,,\qquad \psi=\tilde{\psi}-\frac{\tilde{b}}{\tilde{a}}\tilde{\phi}\,,\qquad t=\tilde{a}g\,\tilde{t}+\frac{1}{g}\tilde{\phi}\,.
\label{d5trans}
\eea
In other words, by substituting \eqref{d5trans} into \eqref{hht} and then dropping the tilde, the metric is identical to the original one. Note that the AdS radius $\ell=1/g$ is invariant under this transformation. While this inversion symmetry that relates the over-rotating metric to an under-rotating one is remarkable, the physical interpretation of the last line in \eqref{d5trans} was not clear.

We note that the metric \eqref{hht} is asymptotically rotating, with angular velocities
\be
\Omega_a^\infty= -a g^2\,,\qquad \Omega_b^\infty=-b g^2\,.
\ee
It is more natural to consider the asymptotically non-rotating frame. This can be achieved by the coordinate transformations $\phi\rightarrow \phi - a g^2\,t$ and $\psi \rightarrow \psi - b g^2\,t$. In this non-rotating frame, the third line of the transformation rules \eqref{d5trans} becomes simply
\be
\phi=g\tilde{t}\,,\qquad \psi=\tilde{\psi}\,,\qquad t=\frac{1}{g}\tilde{\phi}\,.
\ee
In other words, the inversion symmetry is a actually symmetry of interchanging $\phi  \leftrightarrow gt$, while leaving $\psi$ invariant. In the above transformation, we can interchange $a\leftrightarrow b$, together with $\phi\leftrightarrow \psi$. Now the space/time interchanging symmetry becomes
$ \psi \leftrightarrow gt$, while $\phi$ is a singlet.

\section{Space/time symmetry in general dimensions}

We now generalize the five-dimensional space/time interchange symmetry to general dimensions. We first discuss the four-dimensional example, and then give the transformation rules for general $D$ dimensions.

\subsection{$D=4$}

The four-dimensional Kerr-AdS metric was constructed by Carter \cite{Carter:1968ks}. The metric takes the form
\be
ds^2 = -\frac{\Delta_r}{\rho^2}\left(dt-\frac{a}{\Xi}\sin^2\theta d\phi\right)^2+\frac{\rho^2}{\Delta_r}dr^2+\frac{\rho^2}{\Delta _\theta}d\theta^2 + \frac{\Delta_\theta \sin^2\theta}{\rho^2}\left(adt-\frac{r^2+a^2}{\Xi}d\phi\right)^2,
\ee
where
\be
\Delta_r = (1+r^2g^2)(r^2+a ^2)-2mr\,,\quad \Delta_\theta =1-a^2g^2\cos^2\theta\,,\quad \rho^2=r^2 + a^2\cos^2\theta\,,\quad \Xi=1-a^2g^2\,.
\ee
The metric is asymptotically rotating, with the angular velocity $\Omega^\infty=-a g^2$. The metric in the non-rotating frame is
\bea
ds^2 &=& -\frac{\Delta_r}{\rho^2}\left(\frac{\Delta_\theta}{\Xi}dt-\frac{a}{\Xi}\sin^2\theta d\phi\right)^2+\frac{\rho^2}{\Delta_r}dr^2
+\frac{\rho^2}{\Delta _\theta}d\theta^2\cr
&&+ \frac{\Delta_\theta \sin^2\theta}{\rho^2\Xi^2}\Big[a(1+r^2g^2)dt-(r^2+a^2)d\phi\Big]^2.\label{d4nonr}
\eea
We find that it is invariant under the transformation
\bea
m&=&\frac{\tilde{m}}{(\tilde{a} g)^3}\,,\qquad a=\frac{1}{\tilde{a}g^2}\,,\nn\\
r&=&\frac{1}{\tilde{a}g}\tilde{r}\,,\qquad \cos\theta=(1-\tilde{\Xi})^{\frac{1}{2}}\cos\tilde{\theta}\,,\nn\\
\phi&=&g\tilde{t}\,,\qquad t=\frac{1}{g}\tilde{\phi}\,,\label{d4trans}
\eea
and then drop the tilde. This geometric symmetry can also have implications in the physical quantities, such as mass $M$ and angular momentum $J$. They are given by \cite{Gibbons:2004ai}
\be
M=\frac{m}{\Xi^2}\,,\qquad J=\frac{ma}{\Xi^2}\,.
\ee
It is easy to verify that $M$ and $g J$ interchange with each other under the transformation \eqref{d4trans}. This particularly implies that the BPS condition
\be
M + g J=0\,,
\ee
remains invariant under the transformation. However, this clear picture of mass and angular momentum does not extend to higher dimensions.

It is worth observing that the symmetry is also valid when $m=0$, in which case, the metric is AdS vacuum in complicated toroidal coordinates. We can thus obtain the transformation rules in the standard global AdS coordinates. To see this explicitly, we note that if we make the coordinate transformation from $(r,\theta)$ to $(R,\chi)$ via
\be
\fft{(r^2 + a^2) \sin^2\theta}{\Xi} = R^2 \sin^2\chi\,,\qquad
\fft{(1-a^2 g^2\cos^2\theta)(1+r^2 g^2 )}{\Xi} = 1 + R^2 g^2 \,,\label{ato0}
\ee
then the metric \eqref{d4nonr} of $m=0$ becomes the standard AdS$_4$ metric in global coordinates
\be
ds^2 = -(1 + R^2 g^2 ) dt^2 + \fft{dR^2}{1 + R^2g^2 } +
R^2 \left(d\chi^2 + \sin^2\chi\, d\phi^2\right)\,.
\ee
We see that the transformation rules \eqref{d4trans} interchange the left-hand sides of the two equations in \eqref{ato0} up to a ``$-g^2$'' factor, thus leaving the above metric invariant, provided with the interchanging of the space and time as in \eqref{interchange}. The symmetry appears to be rather unremarkable on the AdS$_4$ vacuum, but it is highly nontrivial when the mass parameter does not vanish.

\subsection{General $D$ dimensions}

We now present the transformation rules for the rotating black holes in general $D$ dimensions. The metric in an asymptotically non-rotating frame in Boyer-Lindquist coordinates is \cite{Gibbons:2004uw, Gibbons:2004js}
\bea
ds^2 &=& -W\left(1+r^2g^2\right)dt^2+\frac{2m}{U}\left(Wdt - \sum_{i=1}^{N}\frac{a_i\mu_i^2d\phi_i}{\Xi_i}\right)^2+\sum_{i=1}^{N}
\frac{r^2+a_i^2}{\Xi_i}\mu_i^2d\phi_i^2 \nn\\
&& +\frac{Udr^2}{V-2m}+\sum_{i=1}^{N+\epsilon}\frac{r^2+a_i^2}{\Xi_i}d\mu_i^2
-\frac{g^2}{W(1+r^2g^2)}\left(\sum_{i=1}^{N+\epsilon}
\frac{r^2+a_i^2}{\Xi_i}\mu_id\mu_i\right)^2\,,\label{genmet}
\eea
where
\bea
W= \sum_{i=1}^{N+\epsilon}\frac{\mu_i^2}{\Xi_i}\,, \quad U=r^\epsilon\sum_{i=1}^{N+\epsilon}\frac{\mu_i^2}{r^2+a_i^2}
\prod_{j=1}^{N}(r^2+a_j^2)\,,\nn\\
V=r^{\epsilon-2}(1+r^2g^2)\prod_{i=1}^{N}(r^2+a_i^2)\,,\quad \Xi_i = 1 -a_i^2g^2\,.
\eea
Here, $\epsilon$ is defined by $D=2N+1+\epsilon$. It is 0 for odd $D$ and 1 for even $D$. We also formally set $a_{N+1} = 0$. In this coordinate system, the trigonometric functions $\sin\theta$ and $\cos\theta$ of the latitudinal $\theta$ are generalized to $\mu_i$ which satisfy $\sum_{i=1}^{N+\epsilon}\mu_i^2=1$ and the longitudinal angular $\phi$ and $\psi$ are generalized to $\phi_i$, which are rotating with parameters $a_i$. With these extensions, the transformation rules on the $N+1$ parameters $(m,a_i)$ of the metric are
\bea
m=\frac{\tilde m}{(\tilde{a}_ig)^{D-1}}\,,\qquad a_i=\frac{1}{\tilde{a}_ig^2}\,,\qquad a_j=\frac{\tilde{a}_j}{\tilde{a}_ig}\,, \quad (j\neq i)\,.
\eea
The accompanying coordinate transformations are
\bea
\phi_i&=&g\tilde{t} \,,\qquad t= \frac{1}{g}\tilde{\phi}_i\,,
\qquad \phi_j=\tilde{\phi}_j\,,\quad (j\neq i)\,,\nn\\
r&=& \frac{1}{\tilde{a}_ig}\tilde{r}\,,\qquad
\mu_i^2=\tilde{\mu}_i^2+\sum_{j\neq i}^{N+\epsilon}\frac{\tilde{\Xi}_i}{\tilde \Xi_j}\tilde{\mu}_j^2\,,\qquad \mu_j^2=\left(1-\frac{\tilde{\Xi}_i}{\tilde{\Xi}_j}\right)
\tilde{\mu}_j^2\,,\quad(j\neq i)\,.
\eea
It is easy to verify that $\sum_i \tilde\mu_i^2 = 1$. By substituting all the transformations into \eqref{genmet}, and then dropping the tildes, we find that the metric stays invariant. Note that the unit-length constraint of $\mu_i$ is important in this verification. Furthermore, the transformations are an involution on the individual coordinate: performing them twice maps back to the original coordinate.

Finally, we would like to understand the space/time interchange symmetry in $D=3$, and compare the Kerr-AdS$_3$ to the BTZ black hole \cite{Banados:1992wn}. In $D=3$, the general metric \eqref{genmet} reduces to
\bea
ds^2 &=& - \fft{\Delta_r}{b^2}dt^2 + \fft{r^2 dr^2}{\Delta_r} + \fft{b^2}{\Xi^2} \left(\fft{2ma}{b^2} dt- d\phi \right)^2  \,,\qquad \qquad \Xi=1-a^2 g^2\,,\nn\\
\Delta_r &=& \left(1+g^2 r^2\right)\left(r^2+a^2\right) -2 m r^2\,,\qquad b^2=\Xi\left(r^2+a^2\right)+ 2 a^2 m\,.
\eea
To cast this Kerr-AdS$_3$ metric into the BTZ form, we define $b=\Xi \tilde r$, and then drop the tilde. The metric becomes
\bea
ds^2 &=& -N^2 dt^2 + \fft{dr^2}{N^2} + r^2 \left(N^\phi dt + d\phi\right)^2\,,\nn\\
N^2 &=& 1 + r^2 g^2  - M_{\rm GPP} + \fft{J^2}{4r^2}\,,\qquad N^\phi=-\fft{J}{2r^2}\,,
\eea
where
\be
M_{\rm GPP}=\fft{4 m}{\Xi} \Big(\fft{1}{\Xi}-\fft12\Big)\,,\qquad J=\fft{4 m a}{\Xi^2}\,.
\ee
Here $M_{\rm GPP}$ is the mass defined by \cite{Gibbons:2004ai}, for the asymptotically global AdS$_3$ black hole. Compared to the BTZ black hole, which can be viewed as being asymptotic to planar AdS$_3$, the definition of mass differs only by a numerical constant
\be
M_{\rm BTZ} = M_{\rm GPP} - 1\,.
\ee
The Killing spinors in the BPS limits of the Kerr-AdS$_3$ and BTZ black holes \cite{Coussaert:1993jp,OColgain:2016msw} are more strikingly different \cite{Cvetic:2018ipt}. It is of interest to note that the space/time interchange symmetry now leaves both $M_{\rm GPP}$ and $J$ invariant, and hence it also leaves $M_{\rm BTZ}$ invariant, although the transformation on the $r$ coordinate in the latter case becomes more complicated.

\section{Conclusion}

In this paper, we reviewed the local inversion symmetry found \cite{Chen:2006ea} in the five-dimensional Kerr-AdS black hole constructed by Hawking, Hunter, and Taylor. We improved on this and found that this symmetry is in fact a symmetry of interchanging the time and a spatial longitudinal angular coordinates. We generalized this space/time interchange symmetry to all Kerr-AdS black holes. In four dimensions, this symmetry interchanges the mass and angular momentum and leaves the BPS condition invariant. In three dimensions, the story changes, and the symmetry leaves both the mass and angular momentum invariant. This dimension-dependent property makes it less likely to find a universal simple interpretation on the conserved quantities for general higher dimensions. However, in retrospect, the interchanging symmetry between $t$ and any of the $\phi_i$'s may not be surprising since in the global embedding of AdS$_D$ into flat $(2,D-1)$-space, the time coordinate is also periodic. In fact, $gt$ and the longitudinal $\phi_i$'s all have the same period $2\pi$.

Note that the $N=\lfloor (D-1)/2 \rfloor$ longitudinal coordinates $\phi_i$, latitudinal coordinates $\mu_i^2$ (with $\sum_i \mu_i^2=1$), and rotation parameters $a_i$ form a linear $\mathbb Z_N$ symmetry
\be
a_i\leftrightarrow a_j\,,\qquad \phi_i\leftrightarrow \phi_j\,,\qquad \mu_i^2 \leftrightarrow \mu_j^2\,,
\ee
while the mass parameter $m$ and radial and time coordinates are singlets. By including $\phi_i \leftrightarrow gt$, we now have a $\mathbb Z_{N+1}$ symmetry. Na\"ively one might expect this enlargement will act linearly on $m$ and $a_i$, but actually it acts nonlinearly on these parameters by the inversion of one of the $a_i$'s. Interestingly, it still acts linearly on the coordinates $(r, t, \phi_i, \mu_i^2)$. It should be also pointed out that the manifest involution property of interchanging time and space is nontrivially realized on the accompanying transformations on the parameters and remaining coordinates. For rotating dS spacetimes, the symmetry is better understood in Euclidean signature, where the space is simply the $D$-sphere, with a trivial $\mathbb Z_{N+1}$ symmetry.

It is natural to ask what implication this enlarged symmetry has for the dual conformal field theory (CFT), via the AdS/CFT correspondence. In general higher dimensions, this is not an easy question, since unlike in asymptotically  Minkowski rotating black holes where the parameter $a_i$ is simply the angular momentum per unit mass, such a simple interpretation is lacking in the Kerr-AdS black hole. The AdS/CFT dictionary typically only relates the mass and angular momenta, rather than $(m,a_i)$, of an AdS black hole to the conformal dimension and the spin of the (heavy) dual operator. However, in four dimensions, we saw that the symmetry had the effect of interchanging mass and angular momentum, suggesting a symmetry between the conformal dimension and the spin of the heavy operators in the three-dimensional CFT. In Kerr-AdS$_3$ black holes, the mass and angular momentum are both invariant and hence the symmetry is trivial in the dual field theory. It is of great interest to investigate the implication of our new space/time interchange symmetry for the dual CFT.

\section*{Acknowledgement}

This work is supported in part by the National Natural Science Foundation of China (NSFC) grants No.~11935009 and No.~12375052.

\end{document}